# A STUDY OF PbF$_2$ NANOPARTICLES CRYSTALLIZATION MECHANISM IN MIXED OXYDE-FLUORIDE GLASSES

S. Dyussembekova[a, c], E.E. Trusova[b], S.E. Kichanov[a], D.P. Kozlenko[a]

[a]*Joint Institute for Nuclear Research, 141980, Dubna, Russia*
[b]*Belarusian State Technological University, 220006, Minsk, Belarus*
[c]*Institute of Nuclear Physics, 05003, Almaty, Kazakhstan*

**ABSTRACT**

The heat-treated oxyfluoride glasses with different contents of thulium Tm$^{3+}$ and holmium Ho$^{3+}$ ions have been studied using small-angle X-ray scattering and diffraction. With an increase in relative concentration of added Ho$_2$O$_3$ and Tm$_2$O$_3$ oxides, the growing in the average size of both nanoparticles and local density fluctuations in the glass matrix is observed. In addition, at high relative concentrations of initial rare-earth oxides, the appearance of a crystalline cubic phase PbF$_2$:Tm-Ho is observed, which can provide a source of up-conversion luminescence in the studied glass materials. A change in the concentration of the initial Ho$_2$O$_3$ and Tm$_2$O$_3$ oxides does not effect on a morphology and fractal dimension of the formed luminescence nanoparticles. The structural aspects of nanoparticle formation with up-conversion luminescence phenomenon are discussed.

## 1. Introduction

Currently, transparent glass luminescent ceramics [1-5] are widely used to increase the efficiency of solar cells as near-infrared phosphors, in optical fibers and sensors, in elements for laser technology [1, 6, 7]. First of all, this is due to the formation of non-linear optical properties [2, 8, 9] of glass ceramics, chemical and thermal stability, and ability to control the optical properties and structural characteristics of nanoparticles during their formation [5, 10, 11].

Rare-earth ions are well-known optically active elements, which are sources of effective luminescence, and nanoparticles doped with these ions are characterized by high quantum yields, wide possibilities of tuning optical properties, and noticeable suppression of the effect of concentration quenching. Therefore, luminescent glasses and glass ceramics based on rare-earth ions with a stoichiometric or non-stoichiometric composition are a promising replacement for phosphors single crystals [12]. In this point of view, the up-conversion luminescent glass ceramics are of particular interest [13-15]. The phenomenon of up-conversion luminescence is a joint radiation of a multicomponent system through sequential optical transitions between various optically active ions [13]. Thus, as an example, one can cite the processes of up-conversion luminescence of $PbF_2$ nanoparticles doped with $Yb^{3+}$ and $Eu^{3+}$ ions, in which the emission of infrared radiation is observed during optical pumping of $Eu^{3+}$ ions from the energy levels of $Yb^{3+}$ ions [16, 17].

For glassy nanoceramics, the efficiency of up-conversion luminescence depends not only on the type and concentration of optically active rare-earth ions, but also on the type and composition of the glass matrix [5, 10], the size of the formed nanoparticles [11], annealing modes [5, 18]. It should be noted that an important problem in the development of glassy nanoceramics is the optimization of the glass composition, which effects on the spectral-optical properties of the system. Besides, the stability of the glass materials, when introducing fluorides and oxides of rare earth elements with a molar concentration of few percent [3, 13], is required. Recently, interest in the studies of oxyfluoride and germanium-gallium glasses doped with thulium and holmium ions has grown [19, 20]. These ions have optical transitions in the infrared region with a high energy transfer efficiency of $Tm^{3+} \rightarrow Ho^{3+}$ process during up-conversion luminescence. The presence of holmium ions in glassy nanoceramics makes these materials promising for infrared laser sources with a wavelength of 2 μm [19, 21]. Currently, much attention is paid to optimize the synthesis of such glass ceramics, selecting the optimal ratio of rare-earth ions for the realization of the up-conversion luminescence. Previous studies of germanium-gallium glasses with $Tm^{3+}/Ho^{3+}$ ions indicate a maximum efficiency of up-conversion luminescence to 63% at an initial relative concentration of oxides as $70Tm_2O_3/15Ho_2O_3$ [19]. It is known that the formation of luminescent nanoparticles in silicate composite systems is associated with the chemical processes between oxides of rare earth elements and glass components [10, 18]. However, the structural mechanisms of the luminescent nanoparticle formation in glass matrixes are studied less.

Taking into account the practical aspect of developing glass ceramics based on oxyfluoride glass matrixes, as well as the interest to up-conversion luminescent materials, our work directed at studies the structural aspects of formation of nanoparticles containing thulium $Tm^{3+}$ and holmium $Ho^{3+}$ ions in oxyfluoride glasses using small-angle X-ray scattering and X-ray diffraction methods.

## 2. Materials and methods

The initial glass samples were synthesized by melt-soligification technique in the system $2.2SiO_2$–$1.3GeO_2$–$6.9PbO$–$7.6PbF_2$–$2.0CdF_2$. The introduction of $Ho_2O_3$ and $Tm_2O_3$ oxides in a wide molar range were performed (Table 1). The selected concentration range of $Ho_2O_3$ corresponds to the maximum efficiency of up-conversion luminescence in similar glasses [19].

*Table 1. Concentration of rare-earth oxides*

| Sample label | $Ho_2O_3$ | $Tm_2O_3$ |
|---|---|---|
| *N1* | 0.6 | – |
| *N2* | 0.08 | 0.8 |
| *N3* | 0.4 | 0.8 |
| *N4* | 0.6 | 0.8 |
| *N5* | – | 0.8 |

The raw materials (99.99 % purity) were weighted and homogenized in 20 cm$^3$ aluminum oxide ceramic crucible. The synthesis was performed at maximum temperature 950 °C in an electric furnace for 30 min. The melt was casted to the steel surface and obtained samples were annealed at 300 °C for 4 h in the muffle furnace. Then the glass samples were cooled down to the room temperature with the rate near 50°C per hour. Glass samples were transparent and do not contain cracks, signs or opalescence. The X-ray analysis has confirmed the amorphous nature of the initial glass matrix. The glasses were heat-treated at 350 °C/30 h + 360 °C/50 h in order to form the $PbF_2$ nanoparticles [5].

Small-angle X-ray scattering (SAXS) experiments were performed with a Xeuss 3.0 instrument (Xenocs, France). The radiation was generated by a GeniX3D source (Mo-Kα edge, λ = 0.71078 Å). Small-angle X-ray scattering spectra were obtained using an Eiger2 detector at different sample-detector distances from 1 to 4 m. The thickness of the glass samples was 1 mm. The SAXS spectra were corrected for empty container data. The analysis of small angle scattering data was performed in the software package SasView [22]. X-ray diffraction data were obtained at the detector position at a distance of 0.5 m from the sample.

## 3. Results and discussion

The small-angle X-ray scattering spectra of studied glass materials are shown in Fig. 1. The obtained curves of all samples are similar and have a typical shape for a disordered glass systems [18, 23]. There are small changes in the degree of slope of the curves and their shape, which may correspond to a change in the fractal dimensions of scattering objects. Fig. 1b shows the Guinier plots $\{ln(I(q)), q^2\}$, which is provided the radius of inertia $R_g$ of the scatterers [24, 25]. It can be seen that with an increase in the content of thulium oxide $Tm_2O_3$, there are noticeable changes in the Guinier graphs, which may indicate a clustering or aggregation of smaller particles. On the other hand, the model of several scatters will be correct [5, 18]. In this model, when increase in the content of oxides, then enlargement of the average size of large particles or aggregates is expected. Therefore, to analyze the SAXS data, we used a two-particle model, which postulates the contribution to SAXS curves from luminescent nanoparticles, most likely PbF$_2$:Tm-Ho [18], and from fluctuations density inside the glass matrix [10, 18]. This model has been used previously in studies of other glass systems [5, 18].

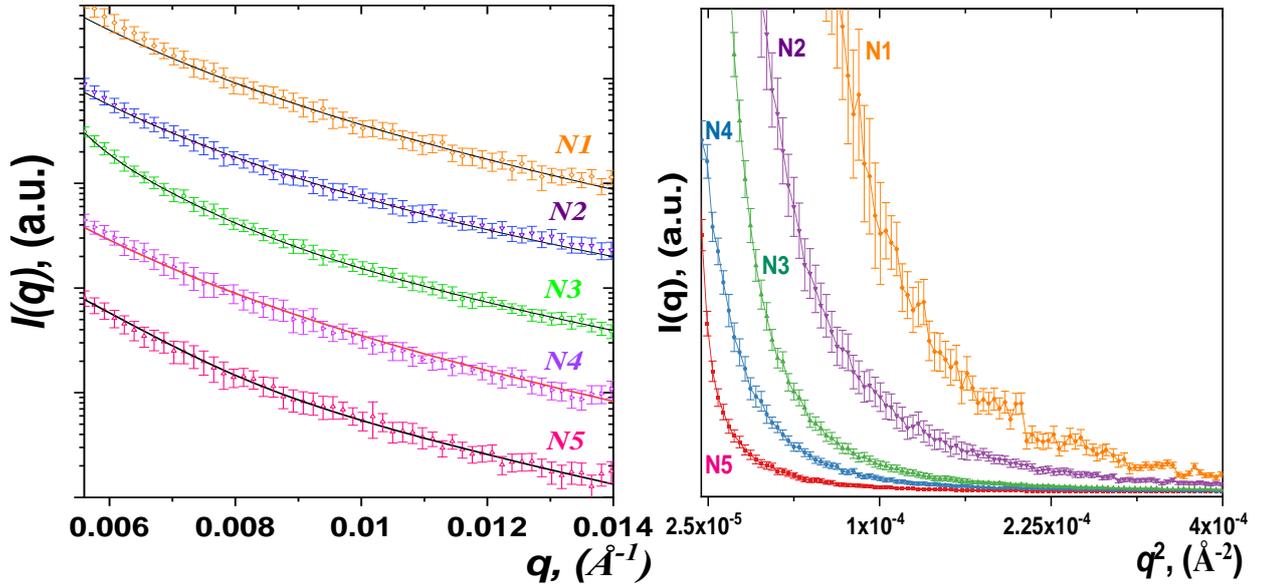

Fig. 1. a) Small-angle X-ray scattering of the studied heat-treated glasses and their approximation by function. b) Guinier plots for SAXS experimental data

Therefore, the obtained SAXS curves were approximated by using the exponential-power law model of Beaucage [26, 27]. It should be noted that even in the absence of nanoparticles in the glass matrix, glass density heterogeneities and defects of different nature can act as scattering objects. The scattering intensity from a system of two scatters is represented by the following expression:

$$I(q) = G_1 \exp\left(\frac{-q^2 R_{g_1}^2}{3}\right) + B_1 \exp\left(\frac{-q^2 R_{g_1}^2}{3}\right)\left(\frac{1}{q^*_1}\right)^{P_1} + G_2 \exp\left(\frac{-q^2 R_{g_2}^2}{3}\right) + B_2 \exp\left(\frac{-q^2 R_{g_2}^2}{3}\right)\left(\frac{1}{q^*_2}\right)^{P_2}, \quad (1)$$

where the coefficients $G_1$, $G_2$, $B_1$ and $B_2$ and the degrees at exponents $P_1$ and $P_2$ are the fitted parameters for the first and second structural level, respectively. The radius of gyration $Rg_1$ and $Rg_2$ correspond to the main parameters of the sizes of scattering objects. The function $q_1^*$ and $q_2^*$ in a power function are normalized as:

$$q_1^* = \frac{q}{\left[\mathrm{erf}\left(\frac{k_1 q R_{g1}}{\sqrt{6}}\right)\right]^3}, \quad q_2^* = \frac{q}{\left[\mathrm{erf}\left(\frac{k_2 q R_{g2}}{\sqrt{6}}\right)\right]^3}, \quad (2)$$

where $k_1$ and $k_2$ are empirical coefficients. The values of the gyration radius $R_{g1}$ and $R_{g2}$ are correlated with fluctuations in the density of glass [10, 18] and with the PbF$_2$:Tm-Ho nanoparticles, respectively.

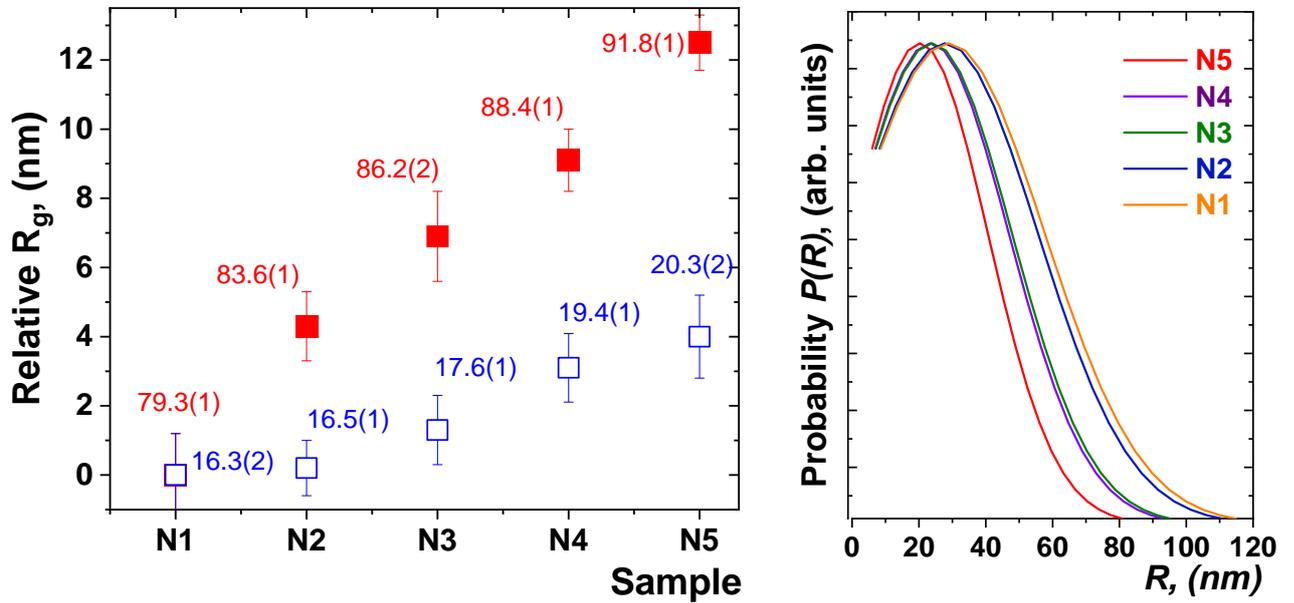

Fig. 2. a) The relative values of the gyration radii $R_{g1}$ and $R_{g2}$ (in nm) calculated from the analysis of the SAXS data for the studied heat-treated glasses. The obtained values are normalized to the corresponding values for N1 sample. b). The distribution of average sizes of nanoparticles in the studied glass samples obtained by equations (3) and (4).

The fractal dimension of scattering objects was estimated and the size distributions of fractal scatterers were determined using classical dependencies for fractal disordered aggregates [28, 29]:

$$I(q) = \phi\, V_{block}(\rho_{block} - \rho_{solvent})^2 P(q)S(q) + B, \quad (3)$$

where $\phi$ is volume fraction of spherical particles in a certain fractal "block" with volume $V_{block}$, $\rho_{block}$, $\rho_{solvent}$ are scattering densities for fractal block unit and solvent, $B$ is scaling background factor, $P(q)$ and $S(q)$ are corresponding scattering laws from separate fractal block and interface reflecting spatial distribution of blocks in the fractal structures [28]. Then the scattering from fractal structure, which consists from blocks is represented:

$$S(q) = 1 + \frac{D_f \Gamma(D_f - 1)}{[1 + 1/q\xi^2]^{(D_f - 1)/2}} \frac{\sin[(D_f - 1)\tan^{-1}(q\xi)]}{(qR_0)^{D_f}}, \quad (4)$$

where $\xi$ is the correlation length corresponded with the block size, $D_f$ is the fractal dimension, $R_0$ is the size of the cluster of blocks.

The results of the approximation of the experimental data by the function (1) and the obtained radius gyration $R_{g1}$ for density fluctuations and $R_{g2}$ for luminescent nanoparticles are shown in Fig. 1a. As it shown in Fig. 2a, the size variation of the density fluctuations and formed nanoparticles presented relatively to the values of N1 sample as $R_g$(Sample)-$R_g$(N1). It can be noted that in heat-treated glasses doped with thulium and holmium ions, the nanoparticles of 42-46 nm in size are formed in a spherical approximation, which the diameter of the nanoparticles is calculated as $D = 2(5/3)^{1/2}R_g$, and the average sizes of density fluctuations in the glass matrix grow from 204(2) nm for sample N1 to 324(3) nm for sample N5 (Fig. 2b). Interestingly, as the relative concentration of $Tm_2O_3$/$Ho_2O_3$ oxides increases, both the average size of nanoparticle clusters and the glass density fluctuations are growing. It can be explained that rare-earth ions are localized not only in nanoparticles, but also in the glass matrix in the form of oxides [5, 18]. In order to estimate the averaged size range of both glass density fluctuations and nanoparticles, approximations of SAXS data using functions (3) and (4) were used. The results of the analysis are shown in Fig. 2b. It can be seen that the average size of nanoparticles in the spherical approximation [28] shifts to the region of large sizes, although the width of the distribution does not change significant. Interestingly, the slope of the SAXS curves practically does not change, and its average value is α= -5.1(5). This indicates that fractal dimensionality and morphology of the nanostructured components of the heat-treated glasses are preserved.

As an important aspect, the structural mechanisms of *$PbF_2$:Tm-Ho* nanoparticles formation can explained also by the detection of crystalline or amorphous state of the luminescent nanoparticles. X-ray diffraction patterns for the studied glassy samples are shown in Fig. 3. All obtained diffraction patterns have a typical shape for scattering from amorphous materials. However, on the X-ray diffraction spectra corresponding to samples N2, N3 and N4, the appearance of several diffraction peaks is observed. This may indicate the formation of a crystal phase from doped thulium and holmium oxides. The positions of the observed diffraction peaks correspond to the cubic structure with the $Fm\bar{3}m$ symmetry and indicate as a $PbF_2$ phase [5]. The unit cell parameter of this phase increases slightly with increasing of the holmium oxide concentration, which indicates the entry of $Ho^{3+}$ ions into the crystal structure of the luminescent nanoparticle. Interestingly, the formation of the crystalline phase is observed only in the simultaneous presence of two oxides $Tm_2O_3$ and $Ho_2O_3$.

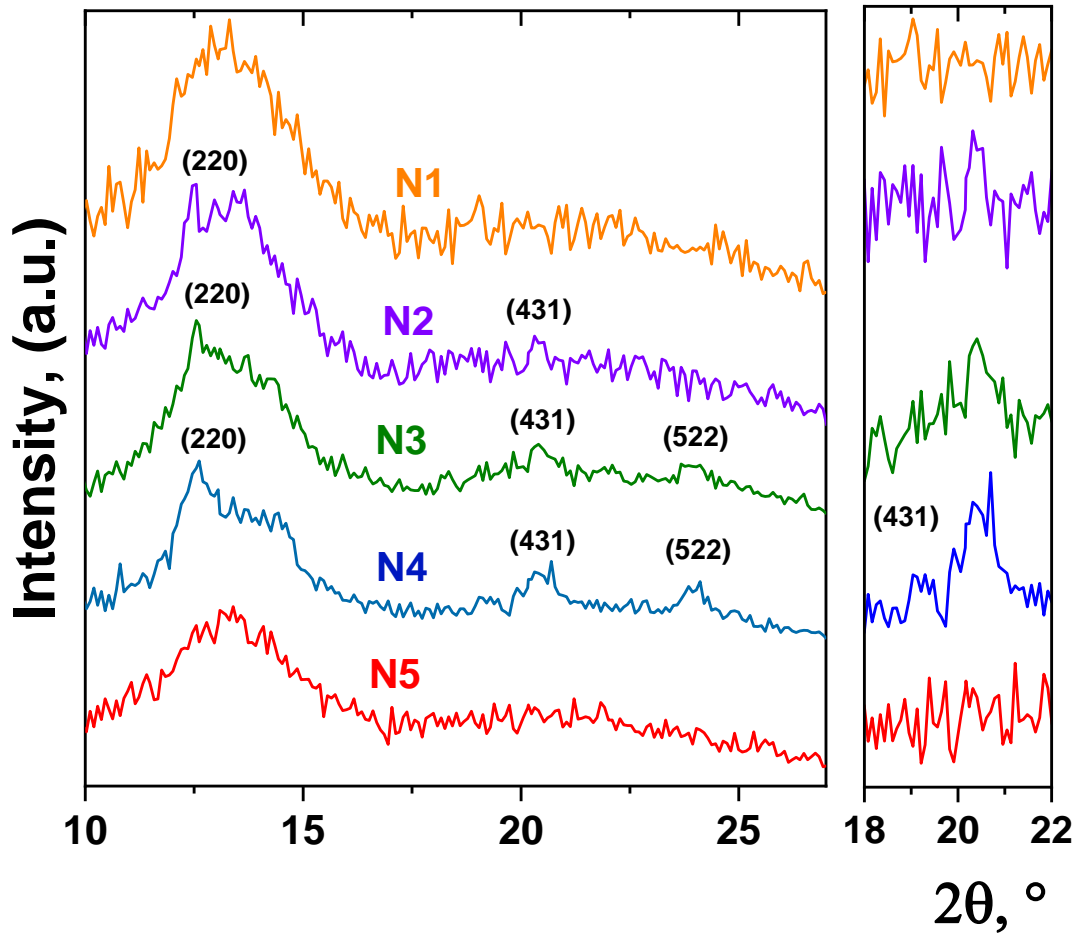

Fig. 3. a) X-ray diffraction patterns of heat-treated glasses. The diffraction peaks of cubic phase $PbF_2$:Tm-Ho are indicated by Miller indices. b) The enlarged section of the diffraction spectra in scattering angle range 18-22°, where the diffraction reflex (431) of the cubic phase $PbF_2$:Tm-Ho is detected.

Based on the obtained experimental data, the following structural mechanism of nanoparticle formation in the heat-treated glass can be proposed. As previously assumed [5, 10, 18], the density fluctuations in the glass materials can serve as the nucleation centers for the oxide nanoparticles $PbF_2$:Tm-Ho. At low concentrations of initial oxides $Tm_2O_3$ and $Ho_2O_3$, the complex amorphous nanostructured structures or aggregates are formed. The nanoparticles form complex branching structures consisting of regular fractal arrangement of clusters inside the glass material. With increasing oxide concentration, the formation of a crystalline phase of $PbF_2$ nanoparticles with changes in the local environment of the glass matrix is observed. These crystallized $PbF_2$ nanoparticles are a host system for rare-earth $Tm^{3+}$ and $Ho^{3+}$ ions, whose entry into the cubic crystal lattice of $PbF_2$, and provide conditions for up-conversion luminescence [5].

## 4. Conclusion

The structural features of nanoparticles formation in heat-treated oxyfluoride glasses have been studied using the small-angle X-ray scattering and X-ray diffraction methods. It has been established that nanoparticles, presumably $PbF_2$:Tm-Ho with sizes of 16-20 nm, are formed at the selected heat-treatment mode. The increase in the average size of the density fluctuations in glass from 71 to 92 nm is observed. With an increase in the concentration of $Ho_2O_3$ and $Tm_2O_3$ oxides, the grow the average sizes of nanoparticles and the sizes of local areas of density fluctuations both was found. The obtained structural information will be useful for the analysis of the optical properties of nanostructured up-conversion-luminescent glass ceramics.